\documentclass[a4paper,11pt,dvitex]{article}
\usepackage{pos}

\title{Canonical partition function and equation of state in the heavy quark region of finite-density lattice gauge theory}
\ShortTitle{Canonical partition function and equation of state in the heavy quark region of FDLGT}

\author*[a]{Shinji Ejiri}
\author[b]{Hideya Hashizume}
\author[b]{Masanari Koiida}
\affiliation[a]{Department of Physics, Niigata University, Niigata 950-2181, Japan}
\affiliation[b]{Graduate School of Science and Technology, Niigata University, Niigata 950-2181, Japan}
\emailAdd{ejiri@muse.sc.niigata-u.ac.jp}

\abstract{We investigate the phase structure and equations of state in lattice gauge theory at finite temperature and finite density, focusing on the canonical partition function obtained by decomposing the grand partition function with respect to particle number.
We demonstrate that when quark masses are heavy, the complex phase of the quark determinant, which causes the sign problem, can be controlled by decomposing the grand partition function according to particle number. 
In this study, we propose an algorithm to calculate the canonical partition function for heavy-quark, high-density effective theory and test it for the first time using one-flavor $SU(2)$ gauge theory, a case where the sign problem arises. 
When the grand partition function of $SU(2)$ is decomposed into canonical partition functions, the sign problem is serious in odd-particle sectors, while it is mild in even-particle sectors. The odd-particle sectors also expose a fundamental issue: center symmetry makes the canonical partition function of $SU(N_c)$ vanish unless the particle number is an integer multiple of $N_c$. 
Although appropriate in the confinement phase, this restriction must be treated carefully in the deconfinement phase. 
We resolve the issue by introducing an infinitesimal external field, calculate the canonical partition functions by Monte Carlo simulation, reconstruct the grand partition function, and determine thermodynamic observables over a wide range of temperature and density.
Furthermore, we find that the chemical-potential dependence of the equation of state at low temperatures exhibits behavior characteristic of the hadron resonance gas model, and that no softening of the QCD matter occurs at finite density.}

\FullConference{The 43rd International Symposium on Lattice Field Theory (Lattice 2026)\\
July 26 to August 1, 2026\\
University of Maryland, College Park, USA}

\begin{document}
\maketitle

\section{Fugacity expansion and center symmetry}

In lattice QCD at finite density, the fermion determinant is a complex number, which prevents the direct application of Monte Carlo methods. 
This makes numerical studies at finite density difficult.
We attempt to calculate the equation of state at finite density by performing the fugacity expansion \cite{Ejiri:2008xt}.
The grand partition function $Z_{\rm GC} (\mu, T)$ is given by the path integral over the gauge fields and is related to the canonical partition function $Z_{\rm C}(N,T)$ at fixed particle number $N$ through the following fugacity expansion:
\begin{equation}
 Z_{\rm GC}(T,\mu)=\int\mathcal{D}U\,(\det M(\kappa,\mu))^{N_{\rm f}}e^{-S_g(U)}, \qquad
 Z_{\rm GC}(T,\mu)=\sum_N Z_{\rm C}(N,T)e^{N\mu/T},
 \label{eq:fugacity}
 \end{equation}
where $e^{\mu/T}$ is the fugacity factor.
When the quark determinant $\det M$ is expanded in terms of the hopping parameter $\kappa$, the resulting terms can be expressed as a sum of products of Wilson loops.
At finite temperature, the temporal direction is compactified with periodic boundary conditions. 
Classifying the terms according to their net winding number $N$ around the temporal direction, we obtain the following fugacity expansion:
\begin{equation}
 \det M(\kappa,\mu)=\sum_N \det M_N(\kappa)e^{N\mu/T}, \qquad
 Z_{\rm C}(N,T) =\int\mathcal{D}U \, ( \det M_N )^{N_{\rm f}}e^{-S_g},  
\label{eq:canonicalpf}
\end{equation}
where we denote the coefficient of $e^{N\mu/T}$ as $\det M_N$.

The Polyakov loop $\Omega$ is the order parameter of center symmetry: 
$\langle\Omega\rangle=0$ in the symmetric phase and $\langle\Omega\rangle\ne0$ in the spontaneously broken phase.
The reason $\Omega$ acts as an order parameter is that it winds around the time direction.
Like $\Omega$, $\det M_N$, which has winding number $N$, transforms nontrivially under the center symmetry.
Under the center transformation, the Polyakov loop and fixed-number determinant transform as
\begin{equation}
 \Omega \longrightarrow \omega \Omega, \qquad \det M_N \longrightarrow \omega^N \det M_N,
 \qquad \omega=e^{2\pi i/N_c}
\end{equation}
for $SU(N_c)$ gauge theory.
Consequently, like $\langle\Omega\rangle$, $Z_{\rm C}(N,T)$ must vanish unless $N$ is a multiple of $N_c$, since
$Z_{\rm C}(N,T)=\omega^N Z_{\rm C}(N,T)$ and $\omega^{N_c}=1$.
Because $Z_{\rm C} (N)$ is the sum of Boltzmann weights for a fixed particle number $N$, this implies that states in which the particle number $N$ is not a multiple of $N_c$ cannot exist.
Furthermore, the complex phase $\omega$ is expected to be one of the origins of the sign problem.

On the other hand, in the deconfinement phase, states in which the particle number is not a multiple of $N_c$ should be realized.
In general, to determine whether a quantity that has a vanishing expectation value because of a symmetry acquires a nonzero value through spontaneous symmetry breaking, one must introduce an infinitesimal external field to break the symmetry, and then take the limit where the volume goes to infinity and the external field goes to zero.
Therefore, for such quantities, the expectation value is determined solely by the response to an external field.
By setting to zero those components that vanish due to the complex phase arising from the center transformation, one may resolve the sign problem associated with that phase. 
Furthermore, the probability of sampling states whose particle number is not a multiple of $N_c$, a key indicator of the deconfinement phase transition, can be calculated by imposing an external field.
It is anticipated that this approach will enable the calculation of the partition function and the equation of state.
In this study, to verify the validity of this method, we calculate the partition function and the equation of state for a simplified version of finite-density $SU(2)$ lattice gauge theory: specifically, one-flavor, heavy-dense $SU(2)$ lattice gauge theory, in which the sign problem arises.

\section{Canonical partition function in $SU(2)$ gauge theory}

From Eq.~(\ref{eq:canonicalpf}), we express the canonical partition function as 
$Z_{\rm C}(N,T) = \langle\det M_N\rangle_{\rm q}Z_{\rm q}$ 
using the partition function and expectation value of quenched QCD, $Z_{\rm q}$ and $\langle \cdots \rangle_{\rm q}$.
For one-flavor $SU(2)$, the $Z_2$ center transformation gives $\det M_N\to(-1)^N\det M_N$ and $\Omega\to-\Omega$.
Thus, the ensemble average cancels exactly for odd $N$, whereas even sectors do not suffer this sign cancellation.

To break the center symmetry explicitly, we add an external source $\epsilon N_s^3 \Omega$
\footnote{A similar discussion regarding the center symmetry and the canonical partition function in $U(1)$ lattice gauge theory is presented in Ref.~\cite{Ejiri:2022jai}.}.
We define 
\begin{equation}
 \langle\det M_N \rangle_{\epsilon}
 =\frac{1}{Z_{\rm q}}\int\mathcal{D}U\,\det M_Ne^{-S_g}e^{\epsilon N_s^3 \Omega}, \qquad
  Z_{\rm C}(N,T) = Z_{\rm q}\langle\det M_N\rangle_{\epsilon}.
\end{equation}
For small $\epsilon$, $e^{\epsilon N_s^3 \Omega}$ can be interpreted as the determinant of an additional heavy fermion.
The lattice size is $N_s^3 \times N_t$.
Averaging over each configuration and its center transform gives
\begin{equation}
 Z_{\rm C}(N,T)=\frac{1}{2} \! \int \! \! \mathcal{D}U \det M_N e^{-S_g}\!
 \left[e^{\epsilon N_s^3 \Omega}+(-1)^N e^{-\epsilon N_s^3 \Omega}\right] =
  \begin{cases}
 Z_{\rm q} \langle\det M_N \cosh(\epsilon N_s^3 \Omega) \rangle_{\rm q},& \text{even},\\
 Z_{\rm q} \langle\det M_N \sinh(\epsilon N_s^3 \Omega) \rangle_{\rm q},& \text{odd}. \nonumber
 \end{cases}
\end{equation}
This averaging procedure makes quantities that must vanish by symmetry exactly zero, even when their direct Monte Carlo calculations have large statistical errors. 
In particular, $Z_{\rm C}(N,T)$ vanishes for odd $N$ in the limit $\epsilon\to0$. 
For small $\epsilon$, the ratios required to reconstruct the grand partition function are independent of $\epsilon$:
\begin{equation}
 \frac{Z_{\rm C}(N+2,T)}{Z_{\rm C}(N,T)}=
 \begin{cases}
 \langle\det M_{N+2}\rangle_{\rm q}/\langle\det M_N\rangle_{\rm q},&N\ \text{even},\\
 \langle\det M_{N+2}\Omega\rangle_{\rm q}/\langle\det M_N\Omega\rangle_{\rm q},&N\ \text{odd}.
 \end{cases}
 \label{eq:ratio}
\end{equation}

\section{Heavy-dense effective theory and numerical method}

To explicitly calculate the canonical partition function, we discuss an effective theory for heavy-quark, high-density $SU(2)$ gauge theory.
The Wilson fermion matrix is
\begin{align}
M_{x,y}={}&\delta_{x,y}-\kappa\sum_{j=1}^{3}\left[(1-\gamma_j)U_{x,j}\delta_{y,x+\hat j}
+(1+\gamma_j)U_{y,j}^{\dagger}\delta_{y,x-\hat j}\right]\nonumber\\
&-\kappa(1-\gamma_4)U_{x,4}e^{\mu a}\delta_{y,x+\hat4}
-\kappa(1+\gamma_4)U_{y,4}^{\dagger}e^{-\mu a}\delta_{y,x-\hat4}.
\label{eq:wilson}
\end{align}
We use the heavy-quark, high-density limit $\kappa\to0$, $e^{\mu a}\to\infty$, with $\kappa e^{\mu a}$ fixed.  Neglecting spatial and backward temporal hopping terms gives heavy-dense QCD \cite{Blum:1995cb,Aarts:2008rr,Aarts:2016qrv,Ejiri:2026ijj}.
The determinant is
\begin{equation}
 \det M =\prod_{\boldsymbol{x}}[1+2C\Omega_{\rm loc}(\boldsymbol{x})+C^2]^2
 \equiv \sum_{n=0}^{4N_s^3} f_n C^n,
 \label{eq:hdqcd}
\end{equation}
where
$C=(2\kappa)^{N_t}e^{\mu/T}$,
$\Omega_{\rm loc}(\boldsymbol{x})=\frac{1}{N_c}\mathrm{tr}\prod_{t=1}^{N_t}U_4(\boldsymbol{x},t)$.
Using $f_n$ in Eq.~(\ref{eq:hdqcd}), 
\begin{equation}
 \det M_n=f_n(2\kappa)^{nN_t},\qquad
 Z_{\rm C}(n)=\langle f_n\rangle_{\rm q}(2\kappa)^{nN_t}Z_{\rm q}.
\end{equation}

The coefficients $f_n$ are obtained as follows.
Assign an index $m = 1, \dots, N_s^3$ to each spatial lattice site, and let $f_n^{(m)}$ denote the coefficient obtained after multiplying by the factors from point 1 up to point $m$.
Using $f_n^{(m-1)}$, $f_n^{(m)}$ can be obtained recursively as follows:
\begin{align}
 f_1^{(m)}={}&f_1^{(m-1)}+4\Omega_{\rm loc}(m)f_0^{(m-1)},\nonumber\\
 f_2^{(m)}={}&f_2^{(m-1)}+4\Omega_{\rm loc}(m)f_1^{(m-1)}
 +[2+4\Omega_{\rm loc}^2(m)]f_0^{(m-1)},\nonumber\\
 f_3^{(m)}={}&f_3^{(m-1)}+4\Omega_{\rm loc}(m)f_2^{(m-1)}
 +[2+4\Omega_{\rm loc}^2(m)]f_1^{(m-1)}
 +4\Omega_{\rm loc}(m)f_0^{(m-1)}.
\end{align}
For $n \geq 4$,
\begin{equation}
 f_n^{(m)}=f_n^{(m-1)}+4\Omega_{\rm loc}(m) f_{n-1}^{(m-1)} +[2+4\Omega_{\rm loc}^2(m)] f_{n-2}^{(m-1)}
 +4\Omega_{\rm loc}(m) f_{n-3}^{(m-1)} +f_{n-4}^{(m-1)}.
 \label{eq:recursion}
\end{equation}
Here, $f_0^{(0)} =f_0^{(m)}=1$ for all $m$ and the initial values are $f_n^{(0)} = 0$ for $n>0$.
These recursive calculations generate all coefficients on every gauge configuration.
After the last point, $f_n=f_n^{(N_s^3)}$.
In particular, $f_0=1$ and
$f_1=4\sum_{\boldsymbol{x}}\Omega_{\rm loc}(\boldsymbol{x})=4N_s^3\Omega$.
The ratio of $Z_{\rm C}(n+2,T)$ to $Z_{\rm C}(n,T)$ is therefore
\begin{equation}
 \ln\frac{Z_{\rm C}(n+2,T)}{Z_{\rm C}(n,T)}=2N_t\ln(2\kappa)+
 \begin{cases}
 \ln \left[ \langle f_{n+2}\rangle_{\rm q} / \langle f_n\rangle_{\rm q} \right],&n\text{ even},\\[3pt]
 \ln \left[ \langle f_{n+2}\Omega\rangle_{\rm q} / \langle f_n\Omega\rangle_{\rm q} \right],&n\text{ odd}.
 \end{cases}
\end{equation}
The canonical functions are obtained by summing these logarithmic ratios from $Z_{\rm C}(0,T)$ for even $N$ and from $Z_{\rm C}(1,T)$ for odd $N$.

\begin{figure}[t]
\vspace{-2mm}
 \centering
 \includegraphics[width=0.31\textwidth]{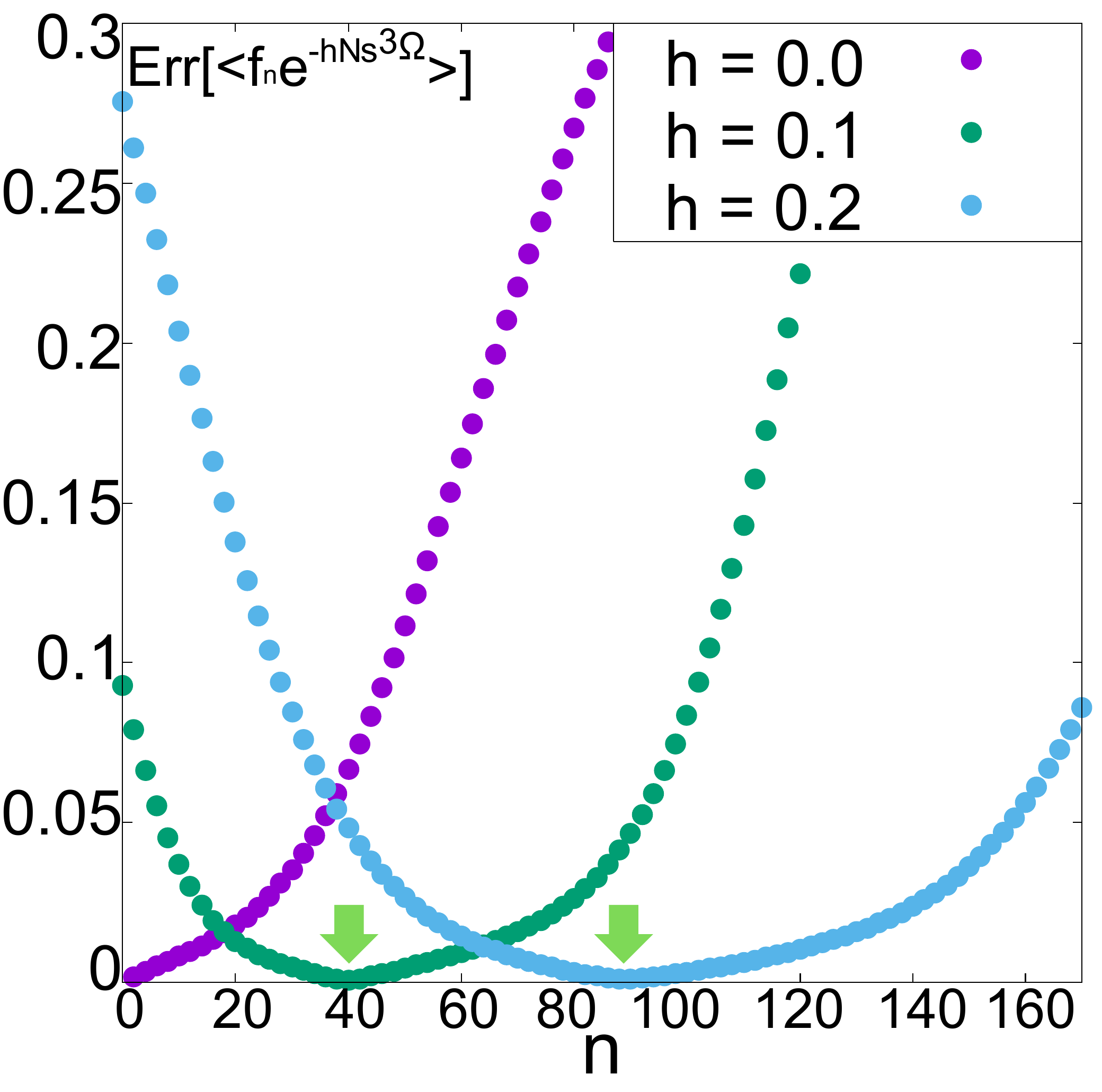}
 \
 \includegraphics[width=0.32\textwidth]{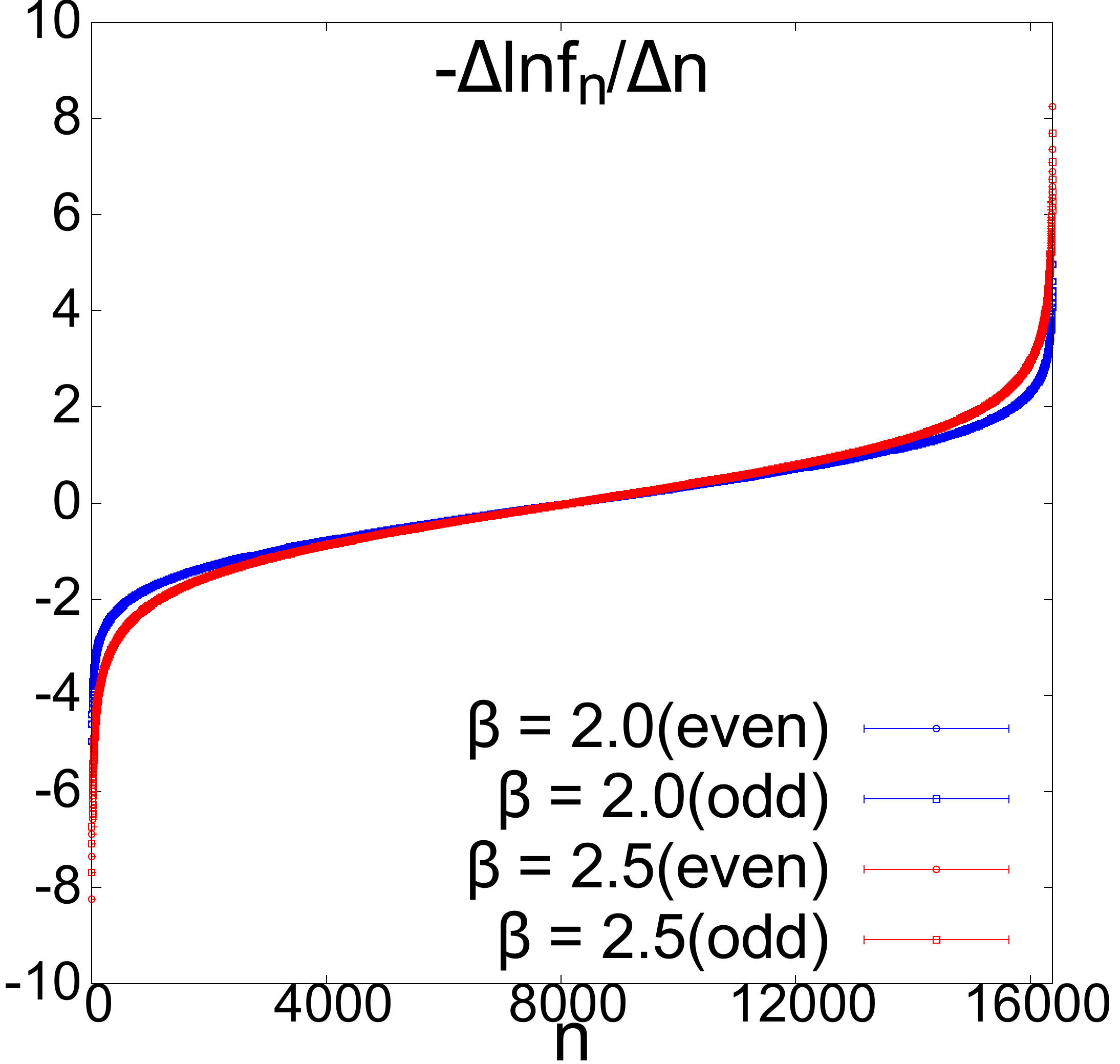}
 \ 
 \includegraphics[width=0.31\textwidth]{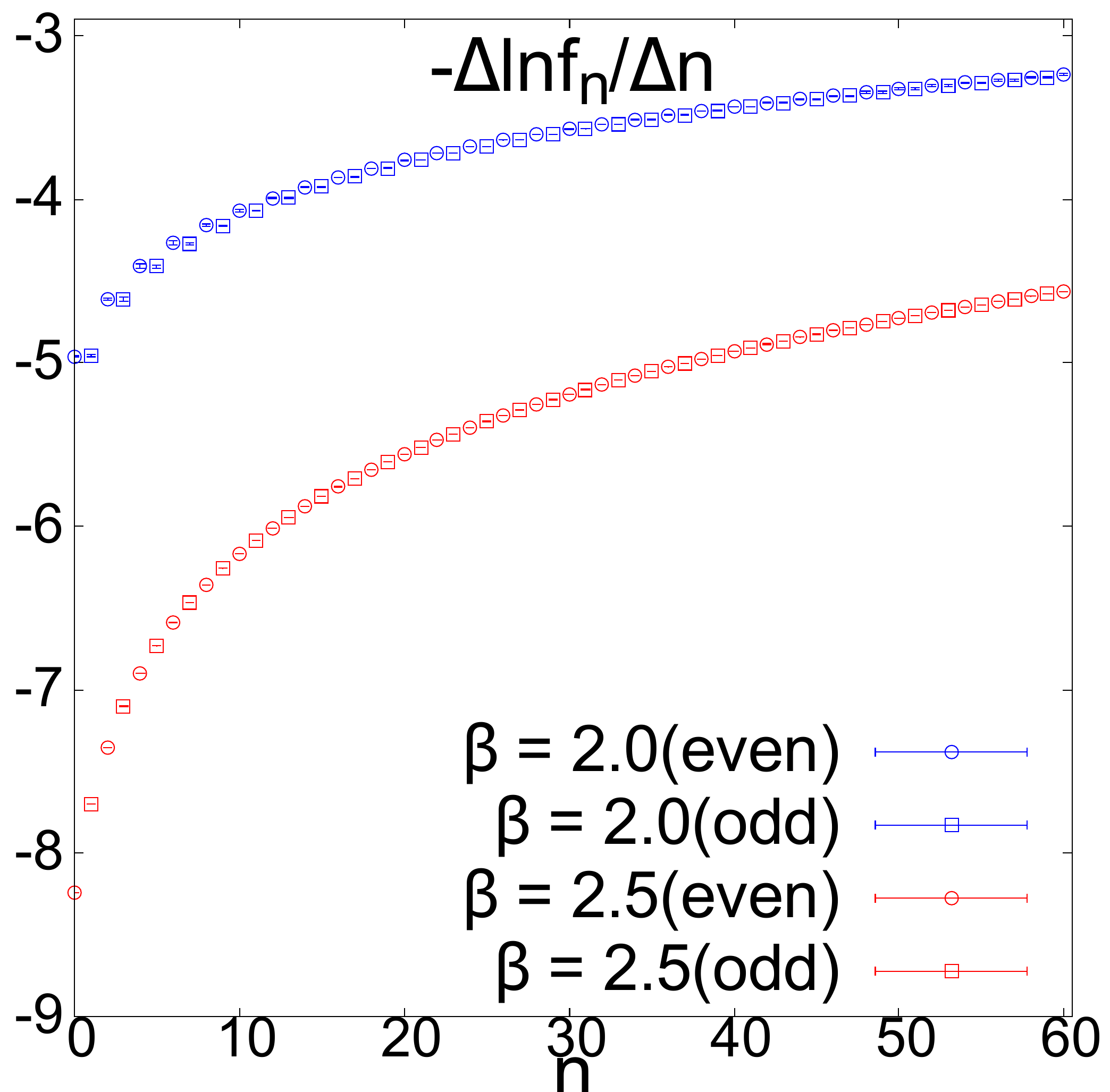}
 \caption{
 Left: Relative statistical error in $\langle f_n e^{-hN_s^3\Omega}\rangle_h$ versus particle number $n$.
 Middle: Discrete derivative of the canonical partition function at $\beta=2.0$ (blue) and 2.5 (red).
 Right: An enlarged view of the low-density region in the middle figure.}
 \label{fig:canonical}
\end{figure}

\paragraph{Overlap problem}
When $n$ is large, an overlap problem arises.
Since approximately $f_n \sim \Omega^n$, configurations near $\Omega=0$ contribute almost nothing to the Monte Carlo statistical sum.
We therefore perform simulations incorporating an external Polyakov loop field, 
\begin{equation}
 \langle O\rangle_h=\frac{1}{Z_h}\int\mathcal{D}U\,O[U]
 e^{-S_g[U]+hN_s^3\Omega[U]},\qquad
 Z_h=\int\mathcal{D}U\,e^{-S_g[U]+hN_s^3\Omega[U]},
\end{equation}
and use the reweighting method,
$ \langle f_n\rangle_{\rm q}=
 \left. \langle f_ne^{-hN_s^3\Omega}\rangle_h \right/ \langle e^{-hN_s^3\Omega}\rangle_h.
 \label{eq:reweight} $
When estimating the optimal source $h$ using a mean-field approximation for the case $f_n \sim \Omega^n$, we obtain $h = N_s^{-3} d(\ln f_n)/d\Omega \approx n N_s^{-3}/\Omega$.
With this value, the condition $\frac{d}{d\Omega}\left[\ln f_n - hN_s^3\Omega\right] = 0$ is satisfied, ensuring that configurations for each $\Omega$ are sampled with equal probability in a Monte Carlo simulation.
Then, the statistical error is minimized.
The left panel of Fig.~\ref{fig:canonical} shows the relative statistical error in $\langle f_n e^{-hN_s^3\Omega}\rangle_h$ as a function of $n$ at $\beta = 2.5$ and $h = 0.0, 0.1, 0.2$ on a $10^3 \times 4$ lattice.
The green arrows represent the optimal values of $n$ obtained by the mean-field approximation.
For each $h$ there is an optimal $n$ with a small statistical error.
We perform simulations for several $h$ values and calculate $f_n$ over a wide range of particle numbers $n$ by combining measurable windows for each $h$.

\section{Canonical partition functions and grand-canonical reconstruction}

The simulations are performed on a $16^3\times4$ lattice. 
The deconfinement transition is at $\beta_c \simeq 2.30$; we compare results at $\beta=2.0$ (confinement) and those at $\beta=2.5$ (deconfinement).
The middle panel of Fig.~\ref{fig:canonical} shows the discrete derivative, 
$-\Delta\ln f_n / \Delta n = -(\ln f_{n+2}-\ln f_n)/2$.
The blue curve represents the results for $\beta=2.0$, and the red curve represents the results for $\beta=2.5$.
The maximum number of particles is $2N_c N_s^3$, which occurs when the lattice is completely filled with particles.
In the simulations, 10,000 configurations are used for each point; calculations are performed while varying the parameter $h$, and the results with the smallest errors are then combined to cover the entire density range.
As seen in this figure, the heavy-dense effective theory possesses high-density/low-density symmetry \cite{Ejiri:2026ijj}. 
The right panel is an enlarged view of the region where $n$ is small.
Circle symbols are the results for even values of $n$, while square symbols are the results for odd values of $n$.
This figure shows that no significant difference is observed between the results for even $n$ and those for odd $n$.

Separating even and odd sectors, the normalized grand partition function can be written as
\begin{equation}
 \frac{Z_{\rm GC}}{Z_{\rm q}} \! = \!
 \sum_{n\,{\rm even}} \! \langle f_n\rangle_{\rm q} \exp\left[n\left(\frac{\mu}{T}+N_t\ln(2\kappa)\right)\right]
 +4N_s^3\langle\Omega\rangle_{\rm q}
 \sum_{n\,{\rm odd}} \! \frac{\langle f_n\Omega\rangle_{\rm q}}
 {\langle f_1\Omega\rangle_{\rm q}} \exp\left[n\left(\frac{\mu}{T}+N_t\ln(2\kappa)\right)\right]. 
 \label{eq:zgc}
\end{equation}
Since $f_0=1$ and $f_1=4N_s^3 \Omega$, the grand partition function can be reconstructed using the data for $ \Delta \ln f_n/\Delta n$.
We note that the odd-particle contribution is proportional to the order parameter $\langle\Omega\rangle_{\rm q}$.
Thus the odd sectors do not contribute in the confinement phase because $\langle\Omega\rangle_{\rm q} =0$, which expresses the absence of a single-fermion contribution, while they survive in the deconfinement phase.

\section{Equation of state and order parameters}

\begin{figure}[t]
\vspace{-2mm}
 \centering
 \includegraphics[width=0.46\textwidth]{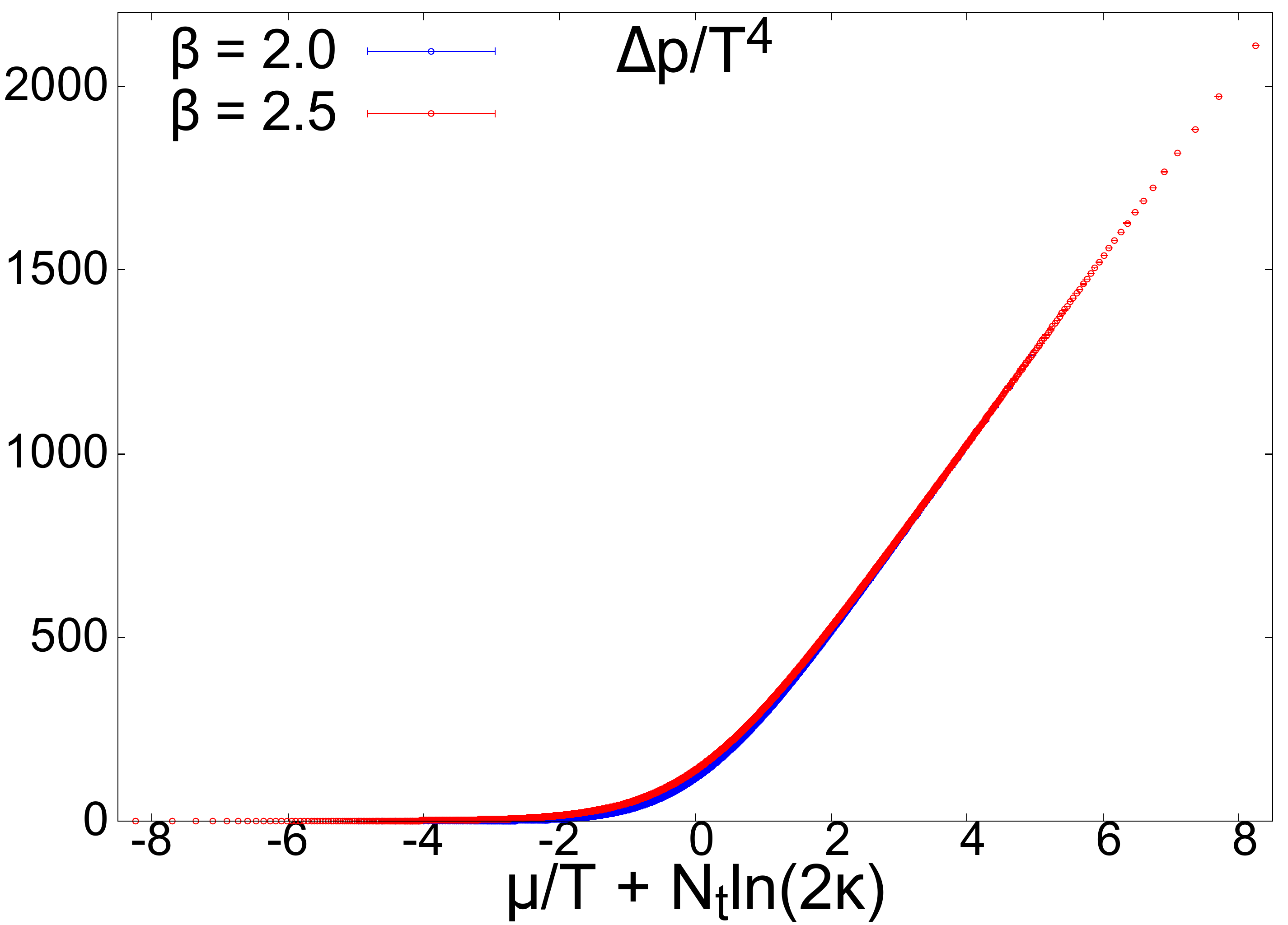}
 \ 
 \includegraphics[width=0.45\textwidth]{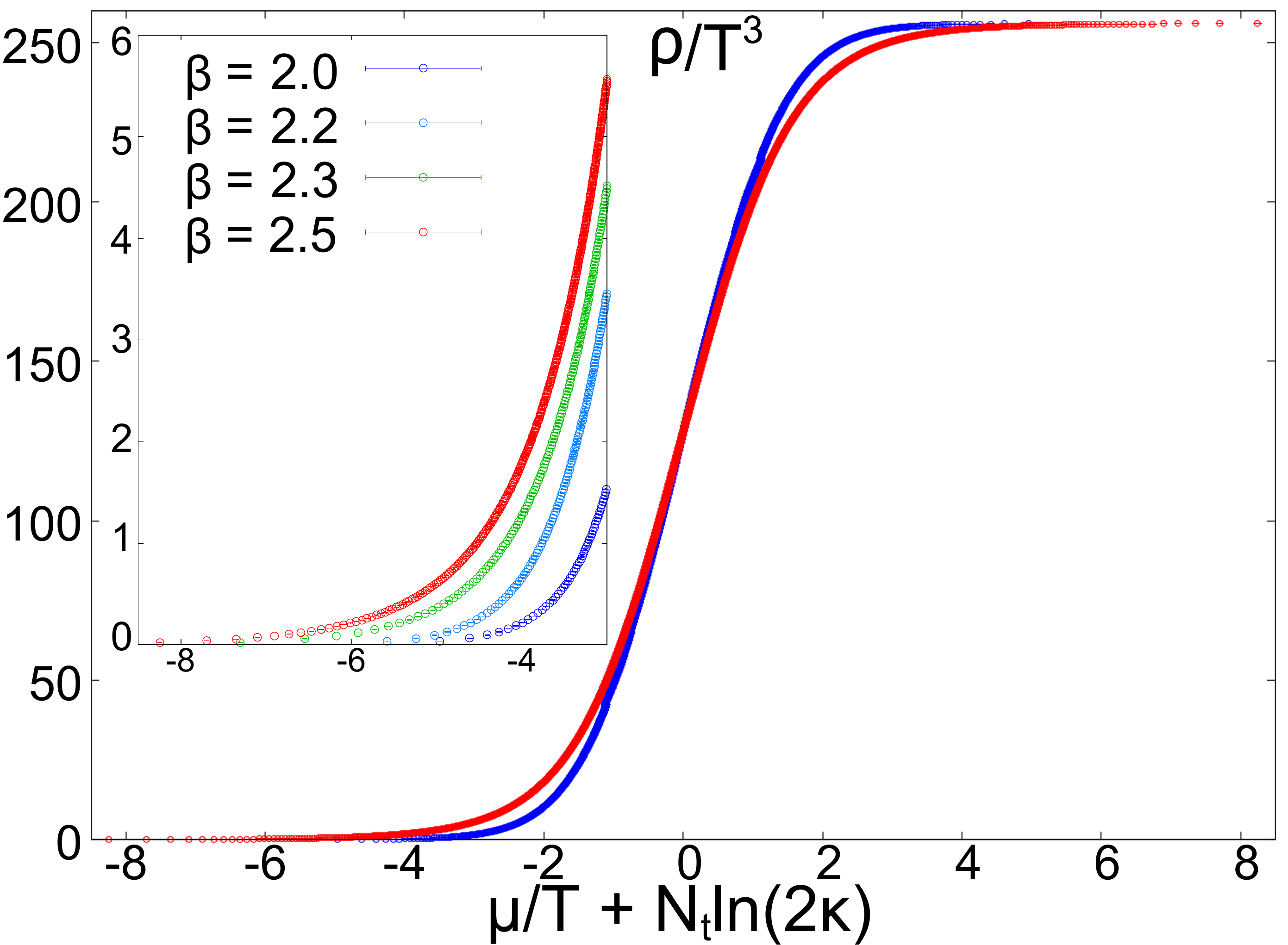}
 \caption{Fermionic contribution to the pressure (left) and particle-number density (right) as a function of $\mu/T+N_t\ln(2\kappa)$ at $\beta=2.0$ and 2.5 on the $16^3\times4$ lattice.}
 \label{fig:eos}
\end{figure}

Pressure and particle-number density are defined by
$p=\frac{T}{V}\ln Z_{\rm GC}$ and 
$\rho=\frac{\partial p}{\partial\mu}=\frac{T}{V}\frac{\partial\ln Z_{\rm GC}}{\partial\mu}$.
The dimensionless density and the fermionic contribution to the pressure are
\begin{equation}
 \frac{\rho}{T^3}=\frac{N_t^3}{N_s^3}\frac{\partial\ln Z_{\rm GC}}{\partial(\mu/T)},
 \qquad
 \frac{\Delta p}{T^4}=\frac{p}{T^4}-\frac{p_{\rm q}}{T^4}=\frac{N_t^3}{N_s^3}\ln\frac{Z_{\rm GC}}{Z_{\rm q}},
 \label{eq:eos}
\end{equation}
where $p_{\rm q}$ is the pressure in the quenched theory. 
More explicitly, differentiating the grand partition function gives
\begin{equation}
\frac{\rho}{T^3}=\frac{N_t^3}{N_s^3}\frac{1}{Z_{\rm GC}/Z_{\rm q}}
\left[ \sum_{n\,{\rm even}}n\langle f_n\rangle_{\rm q} e^{n\left(\frac{\mu}{T}+N_t\ln(2\kappa)\right)}
+4N_s^3\langle\Omega\rangle_{\rm q}
\sum_{n\,{\rm odd}}n\frac{\langle f_n\Omega\rangle_{\rm q}}{
\langle f_1\Omega\rangle_{\rm q}} e^{n\left(\frac{\mu}{T}+N_t\ln(2\kappa)\right)} \right].
\label{eq:densitysum}
\end{equation}
We plot the pressure and the particle-number density in the left and right panels of Fig.~\ref{fig:eos}, respectively.
The blue curve shows the results at $\beta=2.0$ calculated on a $16^3 \times 4$ lattice, while the red curve shows the results at $\beta=2.5$.
The horizontal axis is $\ln C=\mu/T+N_t\ln(2\kappa)$.
As shown in Fig.~\ref{fig:eos}, the density rises rapidly near $\mu/T+N_t\ln(2\kappa)=0$ and approaches fermion saturation at sufficiently large $\mu$.
The change in density in the confinement phase ($\beta=2.0$) is more rapid than that in the deconfinement phase ($\beta=2.5$).
When $\mu/T > -N_t\ln(2\kappa)$, the density saturates and remains constant, while the pressure increases linearly with $\mu/T$ because $\rho=\partial p/\partial\mu$.

In the confinement phase, $\langle\Omega\rangle_{\rm q}=0$ in Eq.~(\ref{eq:zgc}) and only even $n$ contribute.
In $SU(2)$ gauge theory, baryons are composed of two quarks; the absence of any contribution from states with an odd number of quarks is consistent with color confinement.
In the deconfinement phase, on the other hand, $\langle\Omega\rangle_{\rm q}$ takes a nonzero value, and thus contributions from odd $n$ also appear.
This is consistent with the fact that single-quark excitations become allowed in the deconfined phase.
The inset in the right panel of Fig.~\ref{fig:eos} enlarges the low-density region for $\beta=2.0,\,2.2,\,2.3,$ and $2.5$. 
In this region, the density $\rho$ increases exponentially from a value consistent with zero as the chemical potential is increased, even at the larger values of $\beta$.

\begin{figure}[t]
\vspace{-2mm}
 \centering
 \includegraphics[width=0.45\textwidth]{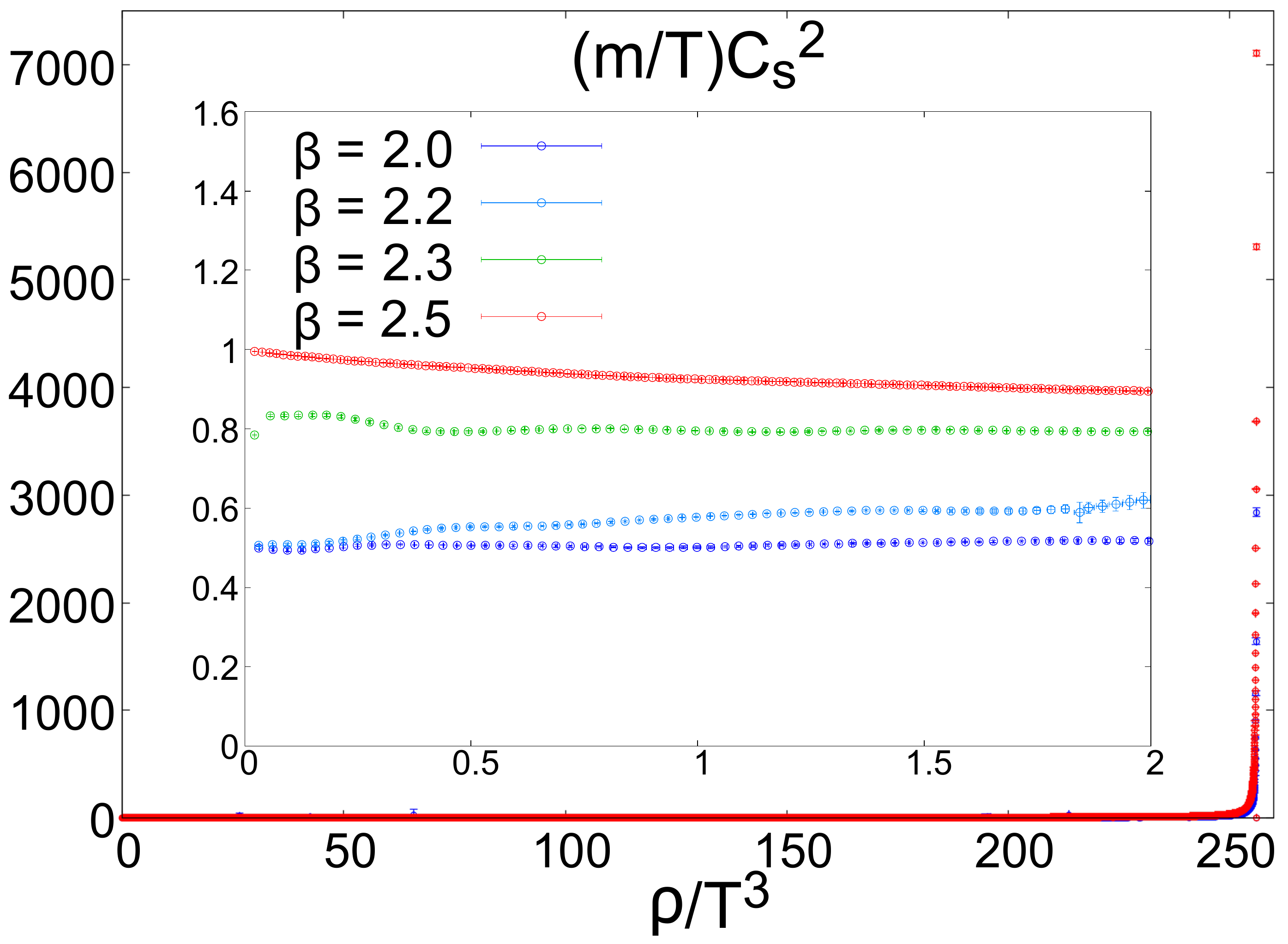}
 \ 
 \includegraphics[width=0.455\textwidth]{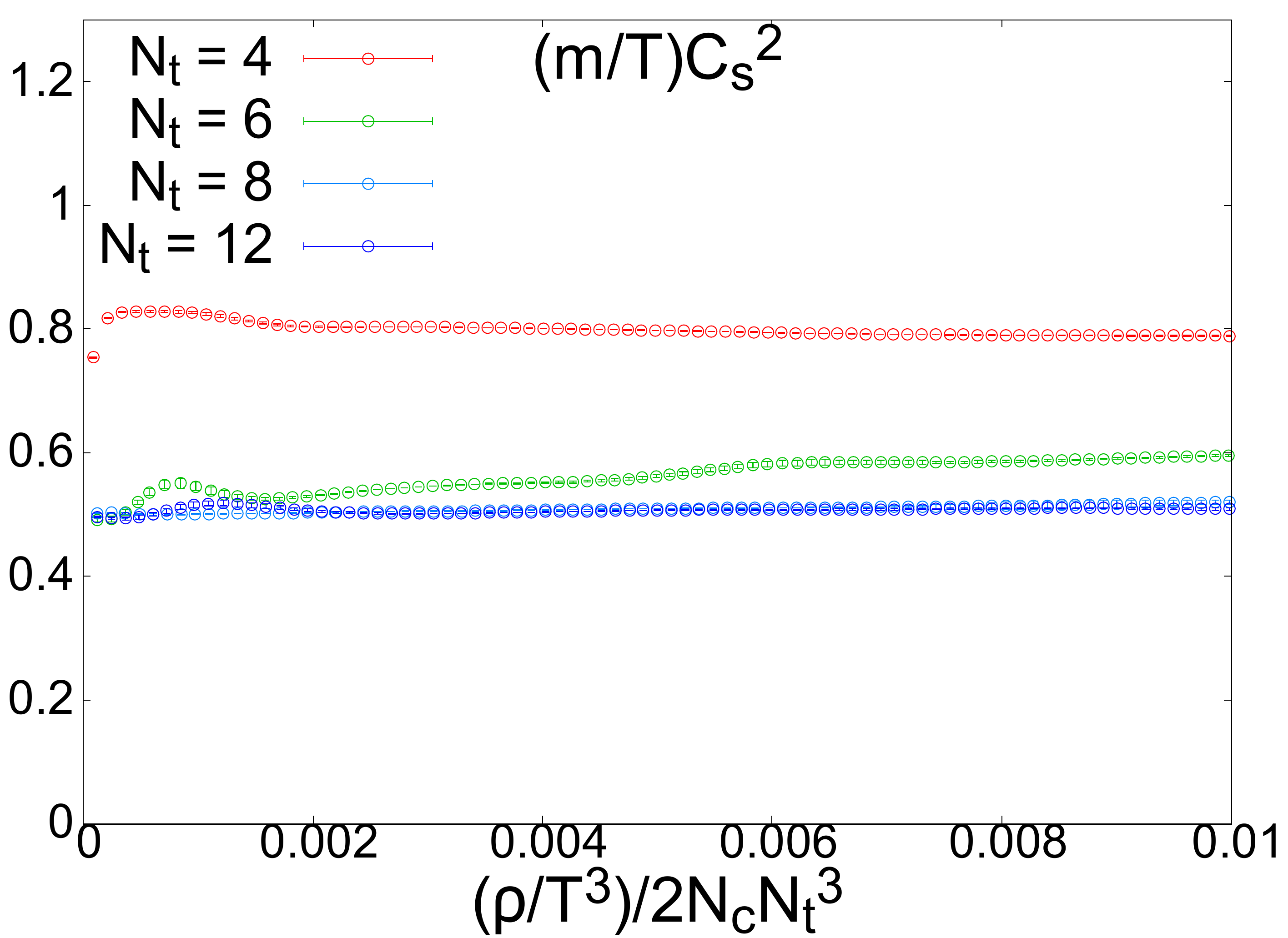}
 \caption{Left: Speed of sound versus $\rho/T^3$ on the $16^3\times4$ lattice 
 at $\beta = 2.0$ (blue), 2.2 (light blue), 2.3 (green), and 2.5 (red).
Right: The results on the $16^3 \times N_t$ lattices with $N_t = 4$ (red), 6 (green), 8 (light blue), and 12 (blue).
The horizontal axis is $(\rho/T^3)/(2N_c N_t^3)= \rho_{\rm lat}/(2N_c)$.}
 \label{fig:sound}
\end{figure}

\paragraph{Speed of sound}
The speed of sound is an important quantity when discussing the softness of the matter.
We use the following nonrelativistic definition:
\begin{equation}
 c_s^2=\frac{dp}{d(m\rho)}=\frac{T}{m}\frac{\partial (\Delta p/T^4)}{\partial (\rho/T^3)}.
 \label{eq:sound}
\end{equation}
Here, $m$ is the effective mass of the particles.
The speed of sound characterizes the stiffness of the matter: a larger $c_s^2$ corresponds to a stiffer equation of state.
This quantity can be calculated from Fig.~\ref{fig:eos} by determining the slope in a plot where $\Delta p/T^4$ is on the vertical axis and $\rho/T^3$ is on the horizontal axis, with $T$ held constant for the partial differentiation.
The results obtained on a $16^3 \times 4$ lattice are shown in the left panel of Fig.~\ref{fig:sound} as the dimensionless quantity $(m/T)c_s^2$. 
Over most of the density range, this quantity is nearly constant, but it diverges near fermion saturation.
The inset is an enlarged view of the low-density region; in the confinement phase, the value is approximately $1/2$ over a wide range of $\rho/T^3$, while in the deconfinement phase, it is approximately unity~\footnote{
In my oral presentation at the Lattice 2026 conference, I stated that $c_s^2$ decreases; however, a bug in the analysis code was discovered immediately after the presentation.
I would like to provide a correction here.
}.
The blue, light blue, green, and red symbols correspond to the results for $\beta = 2.0$, 2.2, 2.3, and 2.5, respectively.
We also performed simulations using $N_t = 4, 6, 8,$ and $12$ at $\beta = 2.30177$, which is the deconfinement transition point for $N_t = 4$~\cite{Ejiri:1998xd}.
These correspond to $T/T_c = 1, 2/3, 1/2,$ and $1/3$, respectively.
We use 100,000 configurations for each simulation.
The results are shown in Fig.~\ref{fig:sound} (right).
In the low-temperature phase the temperature dependence of the speed of sound is small. 

\begin{figure}[t]
 \centering
 \includegraphics[width=0.45\textwidth]{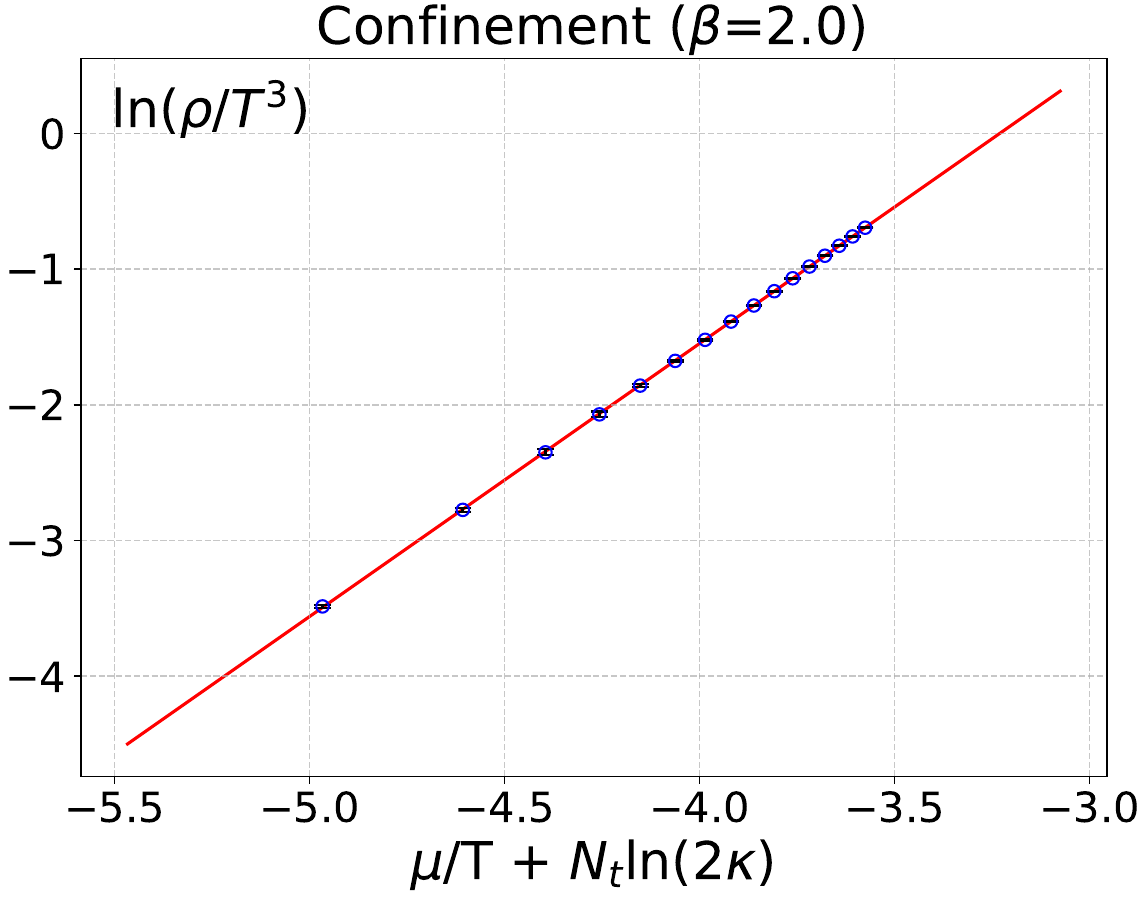}
 \ 
 \includegraphics[width=0.44\textwidth]{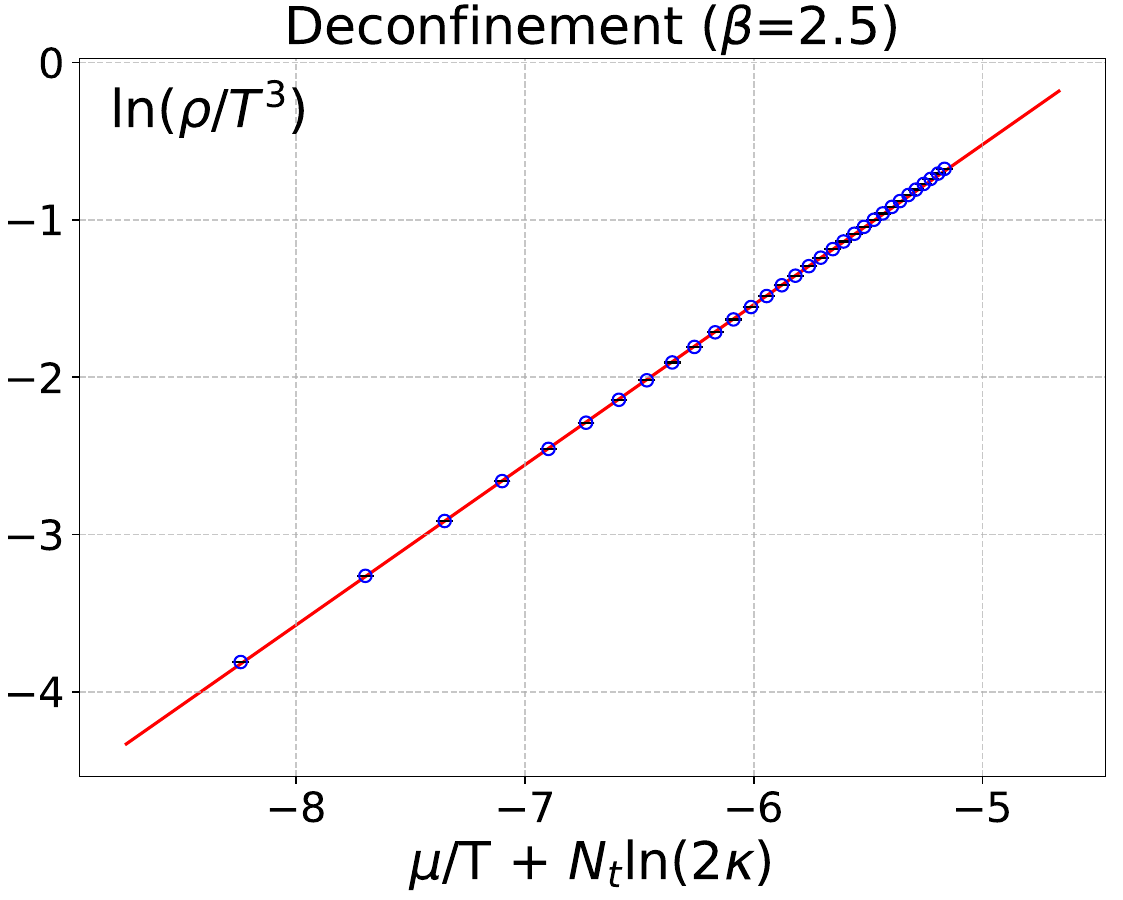}
 \caption{$\ln (\rho/T^3)$ versus $\mu/T$ at $\beta=2.0$ (left) and 2.5 (right).
 The red line shows the fitting result.}
 \label{fig:logrho}
\end{figure}

The result $(m/T)c_s^2=1/2$ in the confined phase can be understood in terms of the hadron resonance gas model.
In the hadron resonance gas model, the pressure is expressed as a sum over meson and baryon states; its dependence on the chemical potential, after subtracting a constant, is proportional to $\cosh(N_c \mu/T)$ \cite{Ejiri:2005wq}.
Here, $N_c$ is the number of colors.
In the heavy-dense effective theory used in this study, the effects of antibaryons are neglected; consequently, the chemical-potential-dependent part of the pressure is proportional to $\exp(N_c \mu/T)$.
In that case, the particle-number density, $\rho = \partial p/\partial\mu$, is proportional to $(N_c/T) \exp(N_c \mu/T)$.
Figure~\ref{fig:logrho} shows the logarithm of the density as a function of $\mu/T$.
The left panel shows the results for the confinement phase at $\beta=2.0$.
We fit the data and determine the slope.
The fitted slope is $2.0119$, in good agreement with the hadron resonance gas prediction $N_c=2$.
On the other hand, the right panel shows the results for the deconfinement phase at $\beta=2.5$, where the slope is 1.0173.
This implies that, at high temperatures, particles consisting of a single quark are present.
Using $\mu$ as the parameter, the speed of sound in the confined phase is
\begin{equation}
 c_s^2 =\frac{1}{m} \frac{\partial (\Delta p)/ \partial \mu}{\partial \rho/ \partial \mu} 
 \approx \frac{1}{m} \frac{(N_c/T) \exp(N_c \mu/T)}{(N_c/T)^2 \exp(N_c \mu/T)}
 = \frac{T}{m} \frac{1}{N_c}.
\end{equation}
These results are consistent with the results at low temperatures shown in Fig.~\ref{fig:sound}.
Thus, the speed of sound does not decrease with increasing density, and no softening of the matter is observed.

\begin{figure}[t]
\vspace{-2mm}
 \centering
 \includegraphics[width=0.45\textwidth]{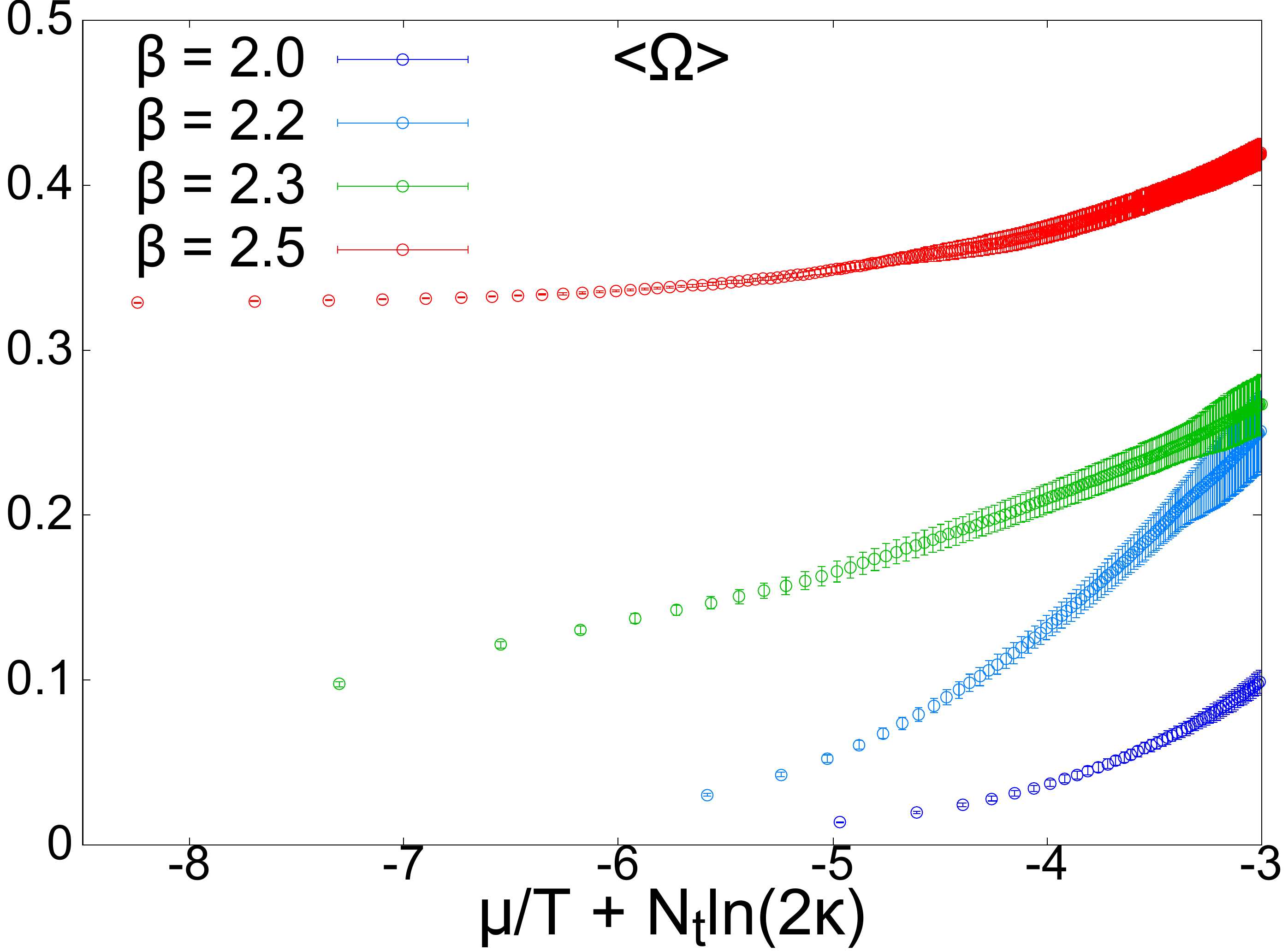}
 \ 
 \includegraphics[width=0.46\textwidth]{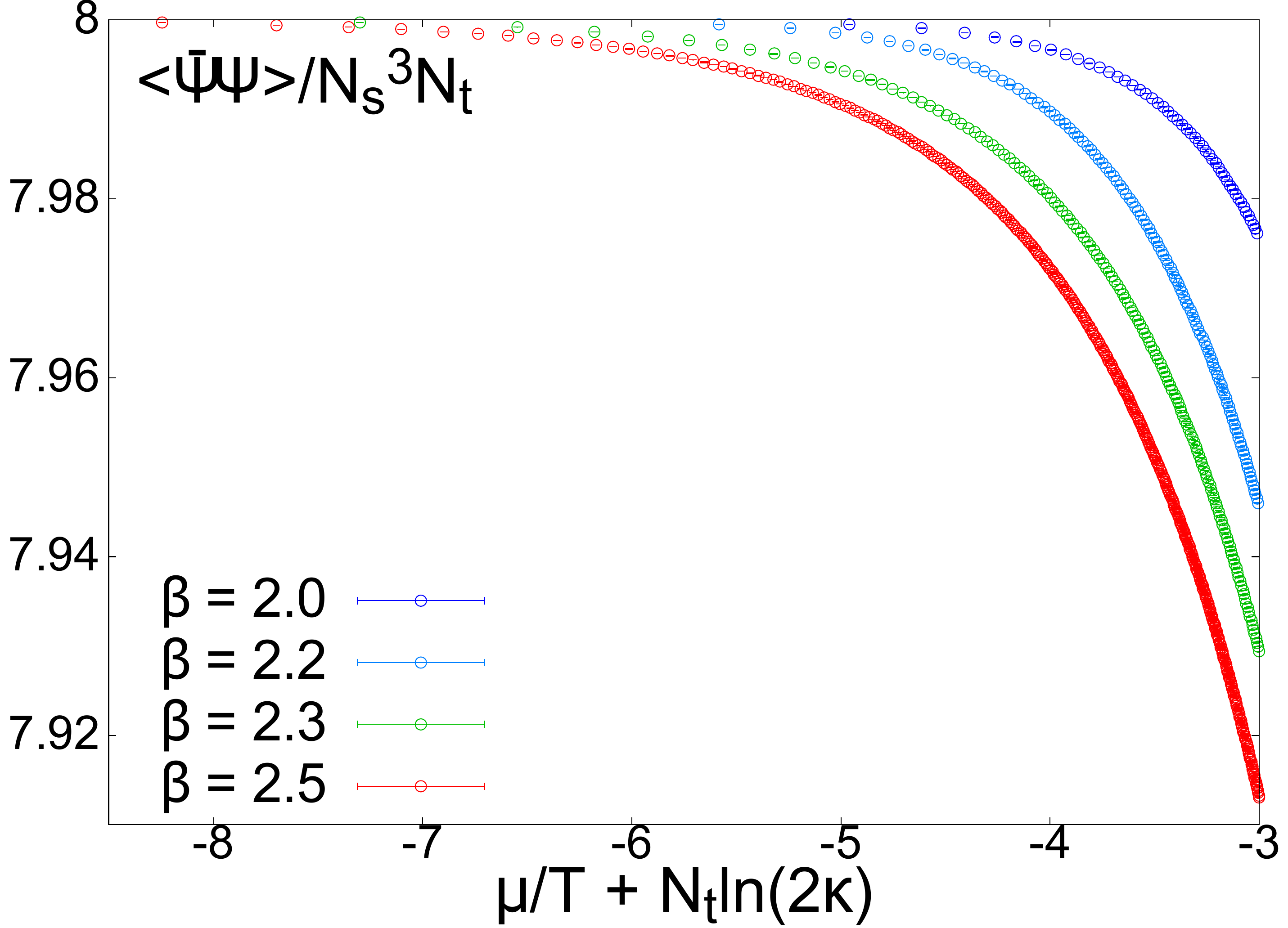}
 \caption{Polyakov loop $\langle\Omega\rangle$ (left) and chiral condensate $\langle\bar\psi\psi\rangle/(N_s^3N_t)$ (right) as functions of $\mu/T$ at $\beta = 2.0$, 2.2, 2.3, and 2.5.}
 \label{fig:chiral}
\end{figure}

\paragraph{Polyakov loop}
Next, we discuss the Polyakov loop, an indicator of confinement.
The Polyakov loop at finite density is reconstructed from the coefficients $f_n$ as follows:
\begin{equation}
\langle\Omega\rangle_{(\mu,T)} =
\frac{\int\mathcal{D}U\, \Omega \det M(\mu)e^{-S_g}}
{\int\mathcal{D}U\,\det M(\mu)e^{-S_g}}
=\frac{
\langle\Omega\rangle_{\rm q}\sum_{n\,{\rm even}}\frac{\langle\Omega^2f_n\rangle_{\rm q}}{\langle\Omega^2\rangle_{\rm q}}e^{nx}
+\langle\Omega f_1\rangle_{\rm q}\sum_{n\,{\rm odd}}\frac{\langle\Omega f_n\rangle_{\rm q}}{\langle\Omega f_1\rangle_{\rm q}}e^{nx}}
{\sum_{n\,{\rm even}} \langle f_n\rangle_{\rm q} e^{nx}
+\langle f_1\rangle_{\rm q}\sum_{n\,{\rm odd}}\frac{\langle \Omega f_n\rangle_{\rm q}}{\langle \Omega f_1\rangle_{\rm q}}e^{nx}},
\label{eq:polyakov}
\end{equation}
where $x=\mu/T+N_t\ln(2\kappa)$ and 
$f_1=4N_s^3\Omega$.
We calculate the Polyakov loop using the $f_n$ data and plot the results in Fig.~\ref{fig:chiral} (left).
When $\mu/T$ is small, $\langle\Omega\rangle$ takes a finite value above the phase transition point, $\beta_c \approx 2.30$, and is zero below it.
At high densities, however, the value becomes finite for all $\beta$.
Even in the region $\beta \geq 2.30$, where the Polyakov loop takes on a finite value, the density remains zero as long as the chemical potential is below the threshold.
It can be seen that $\langle\Omega\rangle$ and $\rho/T^3$ exhibit significantly different behaviors.

\paragraph{Chiral condensate}
Finally, we plot the chiral condensate $\langle\bar\psi\psi\rangle$ at $\beta = 2.0$ (blue), 2.2 (light blue), 2.3 (green), and 2.5 (red) in Fig.~\ref{fig:chiral} (right).
In the heavy-dense effective theory, $\langle\bar\psi\psi\rangle$ is given by 
\begin{equation}
 \langle\bar\psi\psi\rangle=\langle\mathrm{tr}\,M^{-1}\rangle
 =4N_cN_s^3N_t-\frac{Z_{\rm q}}{Z_{\rm GC}}
 \left\langle\kappa\frac{\partial\det M}{\partial\kappa}\right\rangle_{\rm q}
 =N_s^3N_t\left(4N_c-\rho_{\rm lat}\right),
 \label{eq:condensate}
\end{equation}
where $\rho_{\rm lat}$ is the particle number per spatial lattice site in lattice units.
Thus the condensate $\langle\bar\psi\psi\rangle/(N_s^3N_t)$ decreases from $4N_c$ as $\rho_{\rm lat}$ increases.
This behavior is qualitatively similar to the recent tensor-renormalization-group result at $\beta=0$ and $m=0$ \cite{Sugimoto:2025vui}.
However, because the fermions are heavy in this study, the condensate does not vanish.
Since the maximum value of $\rho_{\rm lat}$ is $2N_c$ at saturation, $\langle\bar\psi\psi\rangle/(N_s^3N_t)$ remains finite and approaches $2N_c$.

\section{Conclusions}

Focusing on the canonical partition function, we investigated the phase structure and the equation of state of lattice gauge theory at finite temperature and finite density.
We proposed an algorithm for calculating the canonical partition function in the heavy-dense effective theory and tested it first in one-flavor $SU(2)$ gauge theory.
Due to the center symmetry, the signs in the odd-particle sector alternate, leading to a severe sign problem, whereas no such cancellations occur in the even sector.
The sign problem was resolved by introducing an infinitesimal external field.  
Furthermore, by adjusting this external field, the overlap problem was also resolved.
The reconstructed grand partition function yields the density, pressure, Polyakov loop, sound speed, and chiral condensate across the low- to high-density region. 
In the low-temperature confined phase, the particle-number density is consistent with the hadron resonance gas model. No softening of the equation of state is observed before the system reaches fermion saturation.
Extending this method to $SU(3)$ gauge theory, where three center sectors and a genuinely complex phase must be treated, is the next important step.

\end{document}